\documentclass{elsart}

\usepackage{harvard}

\usepackage{epsfig}

\usepackage{amssymb}

\def\astrobj#1{#1}
\def\url#1{{\ttfamily\def\/{/\discretionary{}{}{}}#1}}
\def\bibcode#1{}

\def\asec {{$\buildrel{\prime\prime}\over .$}}

\begin{document}

\begin{frontmatter}
\title{Stellar Populations and Star Cluster Formation in Interacting
Galaxies with the Advanced Camera for Surveys\thanksref{label1}}

\author[rdg]{Richard de Grijs\thanksref{label2}},
\author[jtl]{Jessica T. Lee},
\author[mcm]{M. Clemencia Mora Herrera},
\author[fva]{Uta Fritze--v. Alvensleben}, and
\author[fva]{Peter Anders}

\address[rdg]{Institute of Astronomy, University of Cambridge, Madingley
Road, Cambridge CB3 0HA, UK}
\address[jtl]{Department of Astronomy, Harvard University, 60 Garden Street,
Cambridge, MA 02138, USA}
\address[mcm]{Departamento de Astronom\'\i a y Astrof\'\i sica, Facultad de
F\'\i sica, Pontif\'\i cia Universidad Cat\'olica de Chile, V. 
Vicu\~{n}a Mackenna 4860 Macul, Santiago, Chile}
\address[fva]{Universit\"atssternwarte, University of G\"ottingen,
Geismarlandstr. 11, 37083 G\"ottingen, Germany}

\thanks[label1]{Based on archival observations with the NASA/ESA {\sl
Hubble Space Telescope}, obtained at the Space Telescope Science
Institute, which is operated by the Association of Universities for
Research in Astronomy (AURA), Inc., under NASA contract NAS 5-26555.}
\thanks[label2]{Corresponding author.\newline
{\it E-mail addresses:} grijs@ast.cam.ac.uk (RdeG),
lee45@fas.harvard.edu (JTL), mcmora@puc.cl (MCMH),
ufritze@uni-sw.gwdg.de (UFvA), panders@uni-sw.gwdg.de (PA)}

\begin{abstract}
Pixel-by-pixel colour-magnitude and colour-colour diagrams -- based on a
subset of the {\sl Hubble Space Telescope} Advanced Camera for Surveys
Early Release Observations -- provide a powerful technique to explore
and deduce the star and star cluster formation histories of the
\astrobj{Mice} and the \astrobj{Tadpole} interacting galaxies.  \\
In each interacting system we find some 40 bright young star clusters
($20 \lesssim {\rm F606W (mag)} \lesssim 25$, with a characteristic mass
of $\sim 3 \times 10^6 M_\odot$), which are spatially coincident with
blue regions of active star formation in their tidal tails and spiral
arms.  We estimate that the main events triggering the formation of
these clusters occurred $\sim (1.5-2.0) \times 10^8$ yr ago.  We show
that star cluster formation is a major mode of star formation in galaxy
interactions, with $\gtrsim 35$\% of the active star formation in
encounters occurring in star clusters.  This is the first time that
young star clusters have been detected along the tidal tails in
interacting galaxies.  \\
The tidal tail of the \astrobj{Tadpole} system is dominated by blue star
forming regions, which occupy some 60\% of the total area covered by the
tail and contribute $\sim 70$\% of the total flux in the F475W filter
(decreasing to $\sim 40$\% in F814W).  The remaining pixels in the tail
have colours consistent with those of the main disk.  The tidally
triggered burst of star formation in \astrobj{the Mice} is of similar
strength in both interacting galaxies, but it has affected only
relatively small, spatially coherent areas. 
\end{abstract}

\end{frontmatter}

\section{Introduction}

Integrated broad-band colours have been used extensively to study the
global star formation histories (SFHs) of elliptical and spiral galaxies
and of spiral bulges.  However, integrated colours are almost always
dominated by the light from the central regions of the galaxies or by
bulge light. 

The use of resolved multi-passband colour data has only recently been
pioneered in efforts to disentangle {\em local} stellar populations from
dust and extinction effects in the disks and bulges of nearby spiral
galaxies \citeaffixed{BP94,T94,dJ96,PB96,P99,E02}{e.g.} and in
intermediate-redshift galaxies in the Hubble Deep Field \cite{F98,A99}. 
Probably the most detailed studies of spatially resolved stellar
populations to date are the recent analysis of M81 by \citeasnoun{K00},
and \citeasnoun{E02}'s ultraviolet--optical colour analysis of {\sl
Hubble Space Telescope (HST)} images of \astrobj{NGC 6753} and
\astrobj{NGC 6782}.  The use of resolved galaxy colours, i.e., the
colours of spatially resolved areas of galactic disks, allows the study
of localised extinction, age distributions and SFHs of spatially
distinct components within a given galaxy, with highly robust results
\citeaffixed{B86,dJ96,A99,P99,E02}{e.g.}. 

\subsection{The Single Stellar Population approximation}

The distribution of colours and spectral energy distributions (SEDs)
across the face of a given galaxy is the cumulative result of the way in
which the local stellar population has been assembled, and of projection
effects. 

In this paper we will, as a first-order approximation, assume that the
integrated {\it local} colours throughout our galaxy images are
sufficiently closely represented by either a single or a superposition
of multiple single stellar populations \citeaffixed{K00}{SSPs;}, i.e.  a
single generation of coeval stars with fixed parameters such as
metallicity, initial mass function (IMF), and internal extinction
\citeaffixed{B97}{see}.  This will allow us to analyse the properties
and SFH of the stellar population {\it that dominates the light} at a
given wavelength.  We note that while this approximation holds
relatively well for young, luminous stellar populations, at older ages
the underlying galactic population is not well represented anymore by a
{\it single} stellar population.  However, as we will show below, for
older ages ($t \gtrsim 10^9$ yr) the age vectors of the individual SSPs
in colour-colour space (defined by the passbands used in this paper) are
almost parallel to each other, so that any deviation from the SSP
approximation will only have a very small effect. 

The use of the relatively well-understood SSPs for the interpretation of
localised colours instead of more complex SFHs simplifies our
interpretation, while the adoption of more complex SFHs changes neither
the results nor the interpretation {\it qualitatively}
\citeaffixed{K00}{see e.g.}. 

In this paper we follow the recent pioneering studies by
\citeasnoun{A99} and \citeasnoun{E02} in constructing colour-colour and
colour-magnitude diagrams (CC, CMDs) on a pixel-by-pixel basis.  This
technique was first used by \citeasnoun{B86}, who used a $B$ vs. 
$(B-R)$ CMD to analyse the stellar populations in \astrobj{NGC 4449}. 
While \citeasnoun{A99} employed the technique to intermediate-redshift
galaxies, in \citeasnoun{E02} we brought it back to the study of stellar
populations in the local Universe.  Here, we explore for the first time
the advantages of using this type of pixel mapping technique in
interacting galaxies.  Galaxy interactions induce violently star-forming
episodes, which are expected to leave signatures in the victims' CC
diagrams and CMDs in the form of distinct features that can be traced
back to, for instance, newly formed stellar populations and their
associated dust lanes. 

Although the local Universe contains a mere handful of interacting
systems at various stages of their gravitational embrace, the addition
of the high-resolution Advanced Camera for Surveys (ACS) to the {\sl
HST} suite of instruments has increased the applicability of
pixel-mapping techniques to interacting systems at greater distances. 
Since the ACS is somewhat undersampled by its point-spread function
(PSF) at optical wavelengths, the individual pixels are statistically
independent. 

In Section \ref{approach.sect}, we give a brief overview of the archival
{\sl HST} images used for our analysis, and of the data reduction
process.  We will then present our results in Section
\ref{results.sect}, which we interpret in the context of star and star
cluster formation induced by the gravitational interactions in Section
\ref{implications.sect}.  For a brief summary of the main results and
conclusions, we refer the reader to Section \ref{summary.sect}. 

\section{ACS observations and reduction: the \astrobj{Tadpole} and
\astrobj{Mice} galaxies}
\label{approach.sect}

Our analysis of the star clusters in and SFHs of the interacting
galaxies \astrobj{UGC 10214} (\astrobj{Arp 188}; \astrobj{VV 029}; the
``\astrobj{Tadpole}'' galaxy) and \astrobj{NGC 4676} (\astrobj{Arp 242};
\astrobj{VV 224}; first nicknamed ``\astrobj{The [Playing] Mice}''
galaxies by \citeasnoun{V58}) is based on the Early Release Observations
(EROs) obtained with the ACS onboard {\sl HST}.  The field of view (FOV)
of these images, taken as part of programme GO-8992 (PI Ford), was
adjusted to optimise the sampling of the tidal tails and debris
surrounding both systems. 

Broad-band images through the F475W, F606W and F814W filters (roughly
corresponding to Johnson-Cousins broad-band {\it B, V} and {\it I}
filters, respectively) of both systems were obtained with the ACS/Wide
Field Camera (WFC), the instrument of choice for red-optimised
optical observations with the {\sl HST}.  The ACS/WFC consists of two
parallel rectangular $2048 \times 4096$ thinned, back-illuminated
Scientific Imaging Technologies (SITe) CCDs, separated by a $\sim 2''$
gap.  With a pixel size of $\sim 0$\asec05, the total FOV is $202''
\times 202''$; this set-up provides a good compromise between adequately
sampling the PSF and a wide FOV \citeaffixed{P01}{e.g.}.  For both
fields, the observations were taken with a relative vertical offset of
$\sim 3''$ to be able to fill in the gap between the two WFC chips. 
The ACCUM imaging mode was used to preserve dynamic range and to
facilitate the removal of cosmic rays. 

The observations of \astrobj{the Mice} were taken on 7 April 2002 (UT);
those of the \astrobj{Tadpole} galaxy were obtained on 1 and 9 April
2002 (UT).  The total exposure times of the observations of \astrobj{the
Mice} used in this paper are 4650s, 4150s, and 3450s in F475W, F606W and
F814W, respectively.  The corresponding total exposure times for the
\astrobj{Tadpole} images are 13780s, 7610s and 8360s. 

We used the ({\sc calacs}) pipeline-processed data products provided by
the {\sl HST} archive, which include the recalibration with new flat
fields (dated 6 August 2002).  These are expected to result in a generic
large-scale photometric uniformity level of the images of $\sim 1$\%. 
In particular, for our analysis we used the final, dithered images
produced by the most recent release of the {\sl PyDrizzle} software
tool.  {\sl PyDrizzle} also performs a geometric correction on all ACS
data, using fourth-order geometric distortion polynomials, and
subsequently combines multiple images into a single output image and
converts the data to units of count rate at the same time. 

We registered the individual images obtained for all passbands and for
both galaxy systems to high (subpixel) accuracy, and created combined
images for each galaxy/passband combination, using the appropriate {\sc
imalign} and {\sc imcombine} routines in {\sc iraf/stsdas}\footnote{The
Image Reduction and Analysis Facility ({\sc iraf}) is distributed by the
National Optical Astronomy Observatories, which is operated by the
Association of Universities for Research in Astronomy, Inc., under
cooperative agreement with the National Science Foundation.  {\sc
stsdas}, the Space Telescope Science Data Analysis System, contains
tasks complementary to the existing {\sc iraf} tasks.  We used Version
2.3 (June 2001) for the data reduction performed in this paper.}. 

\section{The distribution of ages, colours and dust}
\label{results.sect}

\subsection{The brightest star clusters}
\label{clusters.sect}

With the unprecedented combination of resolution and field of view
offered by the ACS, it has now become possible to resolve and study (at
least the bright end of) the star cluster luminosity function (CLF) out
to greater distances than ever before.  In this section, we give the
full observational details for the clusters in the bright wing of the
CLFs of both of our targeted systems. 

The ongoing tidal interactions between both galaxies of \astrobj{the
Mice}, and the tidal disruption of a smaller satellite galaxy by the
dominant Sc-type spiral galaxy in the \astrobj{Tadpole} system have
caused highly variable galactic backgrounds in both systems.  In
addition, the high sensitivity of the ACS, combined with the long
exposure times used for these EROs, reveals large numbers of background
galaxies in the frames of both interacting systems.  Both of these
aspects render the unambiguous detection of genuine star cluster
candidates troublesome. 

We based our initial selection of source candidates on a modified
version of the {\sc daofind} task in the {\sc daophot} software package
\cite{S87}, running under {\sc idl}, in which we selected all potential
sources with peak luminosities -- in any of the passbands -- brighter
than $5 \sigma_{\rm sky}$, i.e.  five times the noise in the sky
background (see Section \ref{systems.sect}).  We required our star
cluster candidates to meet the following stringent selection criteria:

\begin{itemize}

\item Good star cluster candidates must be detected in all of the F475W,
F606W, and F814W passbands.  Upper limits in any of the passbands make a
direct comparison with the underlying galactic stellar populations
impossible, or highly uncertain at best. 

\item Their sizes must be of the same order as the ACS PSF, i.e.,
$\sigma_{\rm cl} \ge \sigma_{\rm PSF}$.  Ground-based test measurements
and artificial PSFs generated by TinyTim \cite{KH01} show that the ACS
PSF is $\sim 0.8 - 1.0$ pixels.  The onboard performance is affected by
optical aberrations and geometric distortions.  Point sources imaged
with the WFC also experience blurring due to charge diffusion into
adjacent pixels because of CCD subpixel variations
\citeaffixed{P01}{e.g.}.  We adopted a conservative size selection
criterion of $\sigma_{\rm cl} \gtrsim 0.8$ pixel. 

\item Genuine star clusters should not be significantly elongated, so
that we limited the roundness parameter in {\sc daofind} to fall in
the range $[-1.0,1.0]$.

\item Finally, the candidate clusters must be associated with the
galaxies, but we do not want to exclude objects that may have been
ejected from the main body of the system due to gravitational kicks. 
Therefore, we required their (conservative) maximum projected distance
to be 2\asec0 and 1\asec5 from the $3 \sigma_{\rm sky}$ contours of the
\astrobj{Mice} and the \astrobj{Tadpole} galaxies, respectively.  At
their respective distances (distance moduli $m-M = 34.93$ and 35.70,
based on their Virgocentric radial velocities and $H_0 = 70$ km s$^{-1}$
Mpc$^{-1}$), this corresponds to a maximum projected distance of $\sim
1$ kpc in both cases. 

\end{itemize}

Finally, we visually verified all of the star cluster candidates by
simultaneously displaying enlargements of them in all three passbands,
and assigned appropriate annuli for aperture photometry to obtain source
and sky background fluxes.  Our final cluster samples contain 45 and 38
sources for \astrobj{the Mice} and the \astrobj{Tadpole}, respectively. 
We used standard aperture radii for the source fluxes of 20 ACS pixels
($\sim$ 1\asec01), and determined the sky background in annular rings
with radii $20 \le r \le 32$ pixels (1\asec01 $\lesssim r \lesssim$
1\asec61), although in a few individual cases these annuli had to be
adjusted to include all of the source flux or avoid bright features in
the background annuli.  Photometric calibration to the STMAG system was
achieved by applying the image header keywords {\sc photflam} and {\sc
photzpt} to the total count rates for each cluster.  As photometric
zero-point offsets we used 25.564, 26.538 and 26.706 for F475W, F606W
and F814W, respectively.  Our bright cluster samples are characterised
by brightnesses of $20 \lesssim {\rm F606W (mag)} \lesssim 25$ in both
systems.  A proper completeness analysis is severely complicated by the
highly variable extinction and (galactic) background levels across the
individual interacting galaxies, and by the small sizes of our objects. 

In Tables \ref{miceclus.tab} and \ref{tadpclus.tab} we list the
clusters' coordinates, properly corrected for the geometric distortions
inherent to the optical design of the ACS, and their photometric
properties in all passbands.  The photometric uncertainties include the
effects of Poisson-type noise, and variations and shot noise in the
background annuli.  Due to the almost parallel age and extinction
vectors, we cannot correct our magnitude measurements for the effects of
extinction (but see Section \ref{implications.sect}).  This clearly
shows, therefore, that the particular combination of F475W, F606W and
F814W passbands by itself is not a good choice for age estimates based
on broad-band colours, for objects older than $\sim 10^8$ yr. 

\begin{table}
\caption[ ]{\label{miceclus.tab}Characteristics of the brightest star
clusters in \astrobj{the Mice}
}
{\scriptsize
\begin{center}
\begin{tabular}{rrrccc}
\hline
\hline
\multicolumn{1}{c}{No.} & \multicolumn{1}{c}{R.A. (J2000)} &
\multicolumn{1}{c}{Dec (J2000)} & \multicolumn{1}{c}{F475W} &
\multicolumn{1}{c}{F606W} & \multicolumn{1}{c}{F814W} \\
& \multicolumn{1}{c}{(hh mm ss.ss)} & \multicolumn{1}{c}{(dd mm ss.ss)}
& \multicolumn{1}{c}{(ST mag)} & \multicolumn{1}{c}{(ST mag)} &
\multicolumn{1}{c}{(ST mag)} \\
\hline
 1 & 12 46 10.35 & 30 45 28.44 & $23.446 \pm 0.064$ & $23.342 \pm 0.026$ & $24.074 \pm 0.037$ \\
 2 &       10.34 &       27.65 & $22.968 \pm 0.033$ & $23.069 \pm 0.022$ & $24.127 \pm 0.040$ \\
 3 &       10.33 &       22.47 & $22.619 \pm 0.029$ & $22.826 \pm 0.023$ & $23.430 \pm 0.027$ \\
 4 &       10.81 &       23.64 & $25.647 \pm 0.181$ & $26.014 \pm 0.173$ & $26.087 \pm 0.159$ \\
 5 &       10.36 &       12.39 & $21.677 \pm 0.021$ & $22.098 \pm 0.020$ & $22.798 \pm 0.025$ \\
 6 &       10.30 &    44 49.12 & $20.658 \pm 0.015$ & $21.196 \pm 0.018$ & $21.794 \pm 0.019$ \\
 7 &       10.86 &       50.57 & $24.819 \pm 0.101$ & $24.717 \pm 0.075$ & $24.653 \pm 0.063$ \\
 8 &       10.29 &       31.03 & $20.338 \pm 0.015$ & $20.733 \pm 0.016$ & $21.348 \pm 0.018$ \\
 9 &       10.19 &       30.32 & $20.779 \pm 0.038$ & $21.141 \pm 0.039$ & $21.707 \pm 0.040$ \\
10 &       10.31 &       30.59 & $20.695 \pm 0.016$ & $21.063 \pm 0.015$ & $21.657 \pm 0.015$ \\
11 &       10.28 &       28.49 & $20.665 \pm 0.019$ & $21.165 \pm 0.021$ & $21.858 \pm 0.023$ \\
12 &       10.30 &       28.46 & $20.732 \pm 0.020$ & $21.198 \pm 0.021$ & $21.900 \pm 0.024$ \\
13 &       10.29 &       27.70 & $20.782 \pm 0.020$ & $21.274 \pm 0.021$ & $21.978 \pm 0.025$ \\
14 &       10.19 &       22.20 & $23.397 \pm 0.294$ & $23.690 \pm 0.276$ & $24.291 \pm 0.301$ \\
15 &       10.22 &       21.35 & $21.325 \pm 0.033$ & $21.938 \pm 0.041$ & $22.385 \pm 0.039$ \\
16 &       10.28 &       17.25 & $21.772 \pm 0.048$ & $22.155 \pm 0.047$ & $22.696 \pm 0.048$ \\
17 &       10.22 &       16.68 & $21.270 \pm 0.022$ & $21.780 \pm 0.025$ & $22.448 \pm 0.030$ \\
18 &       10.28 &       17.31 & $21.737 \pm 0.046$ & $22.117 \pm 0.044$ & $22.673 \pm 0.046$ \\
19 &       10.18 &       09.86 & $21.622 \pm 0.031$ & $22.084 \pm 0.031$ & $22.842 \pm 0.037$ \\
20 &       10.44 &       07.33 & $23.620 \pm 0.077$ & $24.192 \pm 0.120$ & $24.501 \pm 0.141$ \\
21 &       09.96 &    43 59.67 & $20.266 \pm 0.049$ & $20.689 \pm 0.059$ & $21.239 \pm 0.072$ \\
22 &       09.67 &       57.67 & $21.940 \pm 0.067$ & $21.826 \pm 0.063$ & $21.933 \pm 0.064$ \\
23 &       09.92 &       58.44 & $19.372 \pm 0.024$ & $19.735 \pm 0.027$ & $20.323 \pm 0.033$ \\
24 &       10.00 &       51.50 & $20.154 \pm 0.027$ & $20.301 \pm 0.044$ & $19.353 \pm 0.023$ \\
25 &       09.50 &       43.01 & $23.123 \pm 0.077$ & $22.861 \pm 0.072$ & $22.944 \pm 0.080$ \\
26 &       09.98 &       44.36 & $19.678 \pm 0.014$ & $19.770 \pm 0.014$ & $20.595 \pm 0.023$ \\
27 &       09.75 &       26.74 & $23.879 \pm 0.082$ & $23.492 \pm 0.037$ & $23.840 \pm 0.040$ \\
28 &       09.97 &       25.09 & $24.255 \pm 0.095$ & $24.875 \pm 0.125$ & $25.557 \pm 0.197$ \\
29 &       11.79 &       34.79 & $21.428 \pm 0.038$ & $21.848 \pm 0.042$ & $22.753 \pm 0.062$ \\
30 &       11.35 &       31.69 & $22.523 \pm 0.052$ & $23.050 \pm 0.075$ & $24.033 \pm 0.132$ \\
31 &       11.40 &       30.11 & $21.798 \pm 0.022$ & $21.884 \pm 0.022$ & $22.692 \pm 0.050$ \\
32 &       09.59 &       16.66 & $23.547 \pm 0.038$ & $22.950 \pm 0.015$ & $22.107 \pm 0.006$ \\
33 &       12.22 &       27.03 & $23.832 \pm 0.216$ & $24.255 \pm 0.238$ & $24.508 \pm 0.202$ \\
34 &       12.20 &       24.18 & $22.821 \pm 0.043$ & $23.201 \pm 0.044$ & $23.806 \pm 0.050$ \\
35 &       12.16 &       21.43 & $23.261 \pm 0.038$ & $23.576 \pm 0.040$ & $24.319 \pm 0.068$ \\
36 &       11.99 &       19.37 & $21.962 \pm 0.041$ & $22.262 \pm 0.045$ & $22.963 \pm 0.067$ \\
37 &       12.44 &       16.80 & $22.741 \pm 0.033$ & $22.822 \pm 0.028$ & $22.917 \pm 0.023$ \\
38 &       11.58 &       03.80 & $23.756 \pm 0.049$ & $24.331 \pm 0.047$ & $24.797 \pm 0.060$ \\
39 &       10.73 &    42 55.36 & $24.886 \pm 0.110$ & $24.981 \pm 0.070$ & $25.735 \pm 0.121$ \\
40 &       12.00 &    43 02.46 & $22.512 \pm 0.017$ & $23.076 \pm 0.019$ & $23.787 \pm 0.027$ \\
41 &       12.75 &       03.20 & $24.860 \pm 0.100$ & $24.244 \pm 0.031$ & $24.098 \pm 0.022$ \\
42 &       11.77 &    42 56.56 & $24.073 \pm 0.063$ & $24.397 \pm 0.051$ & $25.212 \pm 0.082$ \\
43 &       11.25 &       49.30 & $25.123 \pm 0.137$ & $25.497 \pm 0.106$ & $25.625 \pm 0.099$ \\
44 &       11.58 &       49.95 & $23.949 \pm 0.051$ & $24.464 \pm 0.046$ & $25.107 \pm 0.066$ \\
45 &       12.12 &       47.39 & $25.303 \pm 0.144$ & $25.504 \pm 0.091$ & $25.126 \pm 0.059$ \\
\hline
\end{tabular}
\end{center}
}
\end{table}

\begin{table}
\caption[ ]{\label{tadpclus.tab}Characteristics of the brightest star
clusters in the \astrobj{Tadpole}
}
{\scriptsize
\begin{center}
\begin{tabular}{rrrccc}
\hline
\hline
\multicolumn{1}{c}{No.} & \multicolumn{1}{c}{R.A. (J2000)} &
\multicolumn{1}{c}{Dec (J2000)} & \multicolumn{1}{c}{F475W} &
\multicolumn{1}{c}{F606W} & \multicolumn{1}{c}{F814W} \\
& \multicolumn{1}{c}{(hh mm ss.ss)} & \multicolumn{1}{c}{(dd mm ss.ss)}
& \multicolumn{1}{c}{(ST mag)} & \multicolumn{1}{c}{(ST mag)} &
\multicolumn{1}{c}{(ST mag)} \\
\hline
 1 & 16 06 24.54 & 55 26 06.37 & $24.442 \pm 0.017$ & $24.795 \pm 0.034$ & $25.487 \pm 0.020$ \\
 2 &       23.47 &       07.44 & $25.804 \pm 0.104$ & $25.307 \pm 0.084$ & $26.223 \pm 0.082$ \\
 3 &       21.64 &       14.45 & $24.464 \pm 0.017$ & $24.593 \pm 0.029$ & $25.653 \pm 0.024$ \\
 4 &       22.93 &       03.13 & $25.415 \pm 0.047$ & $25.361 \pm 0.057$ & $26.294 \pm 0.053$ \\
 5 &       22.58 &    25 56.79 & $24.521 \pm 0.024$ & $24.829 \pm 0.047$ & $26.003 \pm 0.043$ \\
 6 &       19.83 &       57.53 & $21.116 \pm 0.015$ & $21.721 \pm 0.017$ & $22.585 \pm 0.021$ \\
 7 &       15.79 &       50.85 & $21.407 \pm 0.006$ & $21.894 \pm 0.008$ & $23.079 \pm 0.010$ \\
 8 &       15.77 &       50.93 & $21.393 \pm 0.006$ & $21.894 \pm 0.008$ & $23.074 \pm 0.010$ \\
 9 &       15.06 &       44.18 & $23.008 \pm 0.010$ & $23.955 \pm 0.029$ & $24.091 \pm 0.013$ \\
 0 &       08.08 &       29.69 & $24.824 \pm 0.078$ & $23.804 \pm 0.029$ & $22.816 \pm 0.007$ \\
11 &       06.96 &       32.37 & $24.932 \pm 0.417$ & $24.704 \pm 0.324$ & $25.748 \pm 0.650$ \\
12 &       06.64 &       32.65 & $21.541 \pm 0.022$ & $21.702 \pm 0.023$ & $21.978 \pm 0.022$ \\
13 &       05.56 &       35.12 & $22.134 \pm 0.055$ & $22.431 \pm 0.063$ & $22.774 \pm 0.059$ \\
14 &       03.57 &       45.37 & $24.977 \pm 0.183$ & $24.540 \pm 0.104$ & $24.563 \pm 0.051$ \\
15 &       07.33 &       19.84 & $25.279 \pm 0.056$ & $24.937 \pm 0.053$ & $26.185 \pm 0.062$ \\
16 &       06.01 &       28.49 & $21.384 \pm 0.036$ & $21.726 \pm 0.041$ & $22.522 \pm 0.059$ \\
17 &       05.47 &       30.49 & $21.651 \pm 0.044$ & $21.796 \pm 0.051$ & $22.722 \pm 0.095$ \\
18 &       02.86 &       38.80 & $22.009 \pm 0.012$ & $22.744 \pm 0.033$ & $25.040 \pm 0.349$ \\
19 &       02.74 &       39.45 & $21.488 \pm 0.012$ & $21.955 \pm 0.022$ & $22.398 \pm 0.031$ \\
10 &       04.59 &       19.52 & $25.078 \pm 0.075$ & $24.323 \pm 0.046$ & $24.935 \pm 0.036$ \\
21 &       03.71 &       21.03 & $24.244 \pm 0.050$ & $24.843 \pm 0.099$ & $25.296 \pm 0.100$ \\
22 &       03.10 &       24.52 & $21.192 \pm 0.012$ & $21.676 \pm 0.025$ & $22.144 \pm 0.035$ \\
23 &       03.71 &       20.91 & $24.500 \pm 0.044$ & $24.867 \pm 0.073$ & $25.253 \pm 0.068$ \\
24 &       01.36 &       34.39 & $22.841 \pm 0.022$ & $22.902 \pm 0.030$ & $23.370 \pm 0.044$ \\
25 &       03.41 &       20.34 & $21.911 \pm 0.011$ & $22.184 \pm 0.015$ & $23.033 \pm 0.019$ \\
26 &       04.56 &       12.53 & $23.703 \pm 0.033$ & $24.178 \pm 0.046$ & $24.744 \pm 0.035$ \\
27 &       02.37 &       26.60 & $21.697 \pm 0.029$ & $21.899 \pm 0.038$ & $22.060 \pm 0.038$ \\
28 &       04.34 &       12.89 & $23.309 \pm 0.030$ & $23.697 \pm 0.037$ & $24.336 \pm 0.031$ \\
29 &       01.17 &       33.31 & $21.673 \pm 0.027$ & $21.583 \pm 0.029$ & $21.981 \pm 0.033$ \\
20 &       02.76 &       20.96 & $21.347 \pm 0.009$ & $21.722 \pm 0.012$ & $22.284 \pm 0.013$ \\
31 &       03.47 &       16.06 & $23.504 \pm 0.019$ & $23.813 \pm 0.029$ & $24.519 \pm 0.023$ \\
32 &       02.75 &       20.84 & $21.357 \pm 0.009$ & $21.721 \pm 0.012$ & $22.282 \pm 0.012$ \\
33 &       00.92 &       32.79 & $21.919 \pm 0.027$ & $21.762 \pm 0.023$ & $22.134 \pm 0.025$ \\
34 &       02.79 &       17.85 & $22.655 \pm 0.020$ & $23.189 \pm 0.027$ & $23.575 \pm 0.022$ \\
35 &       01.35 &       24.59 & $20.563 \pm 0.019$ & $20.852 \pm 0.020$ & $21.492 \pm 0.027$ \\
36 &       01.83 &       19.54 & $20.163 \pm 0.005$ & $20.390 \pm 0.006$ & $20.966 \pm 0.006$ \\
37 &       00.91 &       24.96 & $22.451 \pm 0.033$ & $23.143 \pm 0.052$ & $23.647 \pm 0.064$ \\
38 &       01.92 &       16.63 & $22.821 \pm 0.021$ & $22.891 \pm 0.022$ & $23.381 \pm 0.017$ \\
\hline
\end{tabular}
\end{center}
}
\end{table}

\subsection{Properties of the interacting systems as a whole}
\label{systems.sect}

The analysis of broad-band galaxy colours in terms of their SFHs and
stellar content is complicated due to the composite nature of the
galaxy's stellar population, the age-metallicity degeneracy, the effects
of internal extinction, possible variations in the initial mass function
\citeaffixed{A99}{IMF; e.g.}, and the uncertainties in the absolute age
zero point arising from these.  Therefore, the use of spatially resolved
colours has major advantages over that of the integrated colours alone. 

\begin{figure}
\psfig{figure=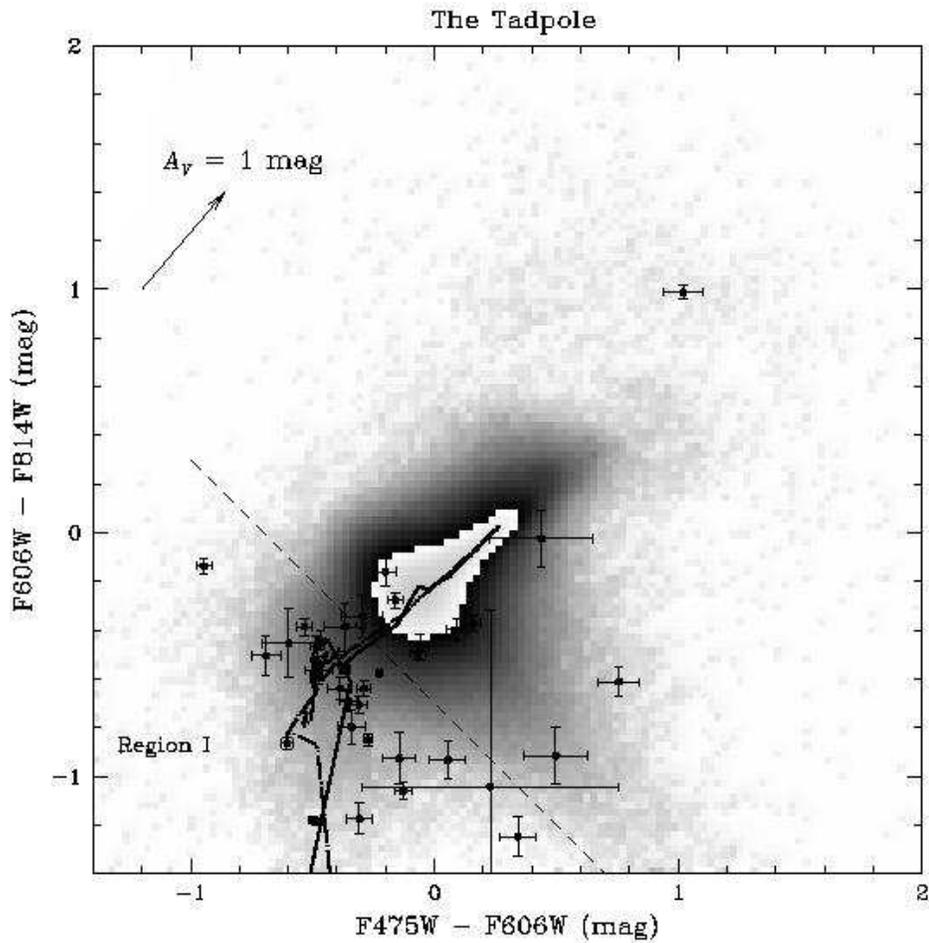,width=14cm}
\caption{\label{cctadp.fig}Pixel CC diagram for the \astrobj{Tadpole}
galaxy.  Region I is demarcated by the dashed line.  The bullets with
their associated error bars represent the cluster colours and their
uncertainties; the thick solid and dash-dotted lines are the model
predictions for the evolution of SSPs with solar metallicity ($Z =
0.020$) and $Z = 0.008$, respectively.  The features in the models near
$({\rm F475W}-{\rm F606W}) = -0.6$ and $({\rm F606W}-{\rm F814W}) =
-0.4$ are caused by the appearance of red supergiants.}
\end{figure}

\begin{figure}
\psfig{figure=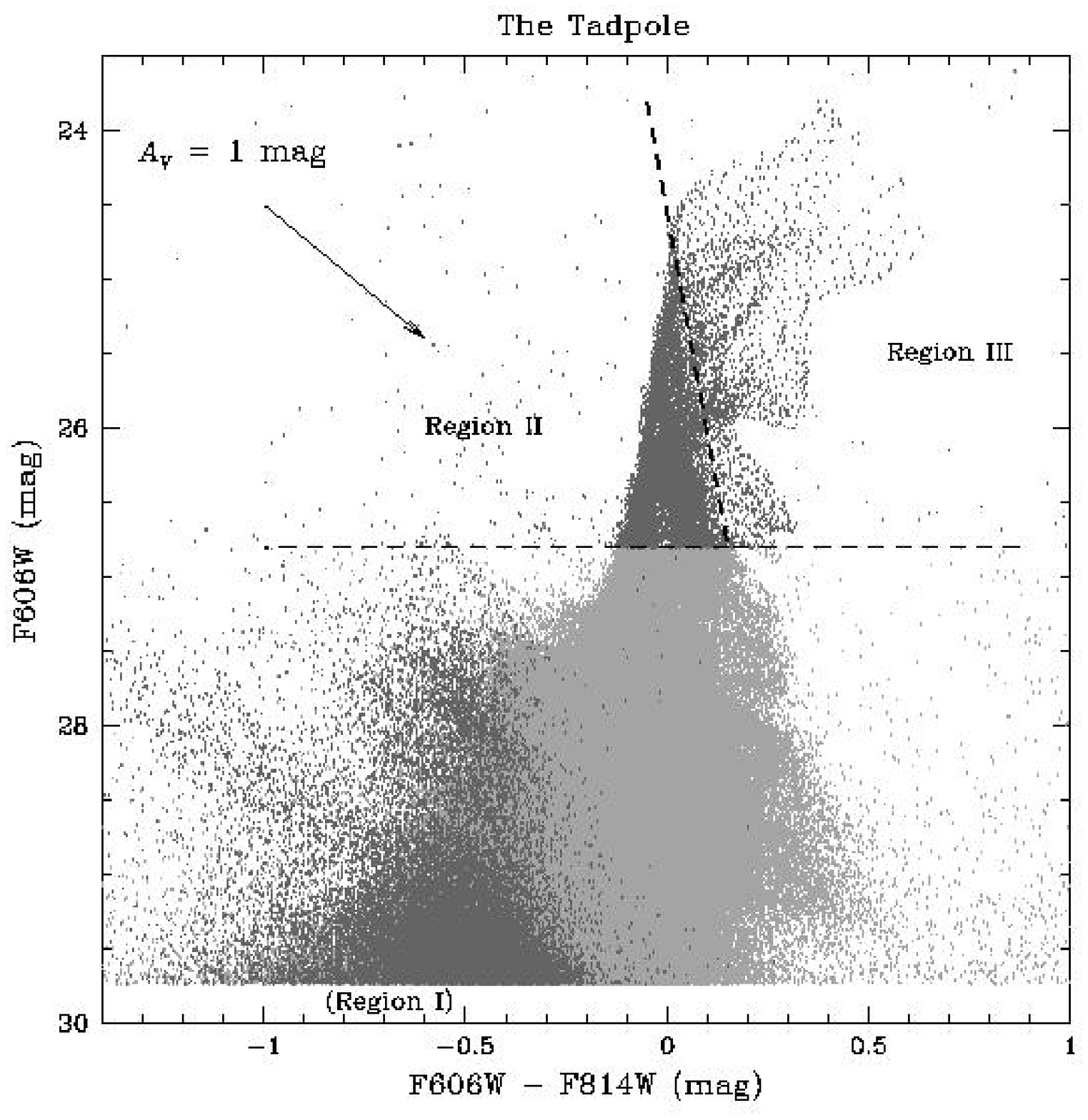,width=14cm}
\caption{\label{cmdtadp.fig}Pixel CMD for the \astrobj{Tadpole} galaxy. 
Specific features in the CMD are indicated by darker shading of the
corresponding pixels and demarcated by dashed lines.  The label
``(Region I)'' indicates the pixels corresponding to Region I defined on
the basis of the CC diagram of Fig.  \ref{cctadp.fig}.}
\end{figure}

We constructed CC diagrams and CMDs for both the \astrobj{Tadpole} and
the \astrobj{Mice} galaxies on a pixel-by-pixel basis, following the
recent studies by \citeasnoun{A99} and \citeasnoun{E02} that showed the
viability of such an approach.  To avoid the spurious effects of noise
in the sky background and to optimise the scientific return with respect
to the processing time of the millions of pixel values, we only consider
pixels with a minimum count level in the image with the lowest
signal-to-noise (S/N) ratio.  For the \astrobj{Tadpole} system, this
translated to a minimum count level of three times the sky noise in the
F606W image, and for the \astrobj{Mice} galaxies we used a $4
\sigma_{\rm sky}$ level in the F475W image.  The pixel CC diagrams,
$({\rm F475W}-{\rm F606W})$ vs.  $({\rm F606W}-{\rm F814W})$, and CMDs,
F606W vs.  $({\rm F606W}-{\rm F814W})$, of both galaxies are shown in
Figs.  \ref{cctadp.fig}, \ref{cmdtadp.fig}, \ref{ccmice.fig} and
\ref{cmdmice.fig}.  The data have been corrected for foreground
extinction using the \citeasnoun{S98} extinction map and the Galactic
reddening law of \citeasnoun{RL85}: $A_{\rm F475W}/A_V \simeq 1.24;
A_{\rm F606W}/A_V \simeq 0.90; A_{\rm F814W}/A_V \simeq 0.50$ mag.  The
arrows indicate the reddening and fading vectors for $A_V = 1.0$ mag
(internal) extinction, assuming a foreground screen dust geometry. 

\subsubsection{Dissecting the \astrobj{Tadpole}}

We have also included our cluster colour measurements in the CC diagrams
of Figs.  \ref{cctadp.fig} and \ref{ccmice.fig}.  Close inspection of
the density distribution of pixels in the CC diagram of the
\astrobj{Tadpole} (Fig.  \ref{cctadp.fig}) reveals a sharp drop in
density to bluer colours below the dashed line.  The lower-density
plateau bluewards of this drop coincides approximately with the loci of
the majority of the star clusters.  Therefore, we define the region
characterised by $({\rm F606W}-{\rm F814W}) \le -0.7 \le ({\rm
F475W}-{\rm F606W})$ as \astrobj{Tadpole} Region I of special interest. 
The pixels in this region also occupy a coherent region in
colour-magnitude space, although not as sharply delineated as in
colour-colour space (see Fig.  \ref{cmdtadp.fig}).  The slight density
enhancement in Fig.  \ref{cctadp.fig} towards redder colours in $({\rm
F475W}-{\rm F606W})$ and bluer colours in $({\rm F606W}-{\rm F814W})$ is
an artificial feature caused by the lowest-S/N pixels in the F606W
passband, located at the interface region between the galaxy and the sky
background. 

In the \astrobj{Tadpole} CMD of Fig.  \ref{cmdtadp.fig}, we define two
further regions of interest.  Region II consists of the pixels in the
well-defined, bright, sharply delineated peak at F606W $\le 26.8$ mag;
Region III is the fuzzy swarm of pixels on the redward side of that same
peak.  These three features occupy similar well-defined loci in all of
the other CMDs that can be constructed from the available observations. 

\begin{figure}
\psfig{figure=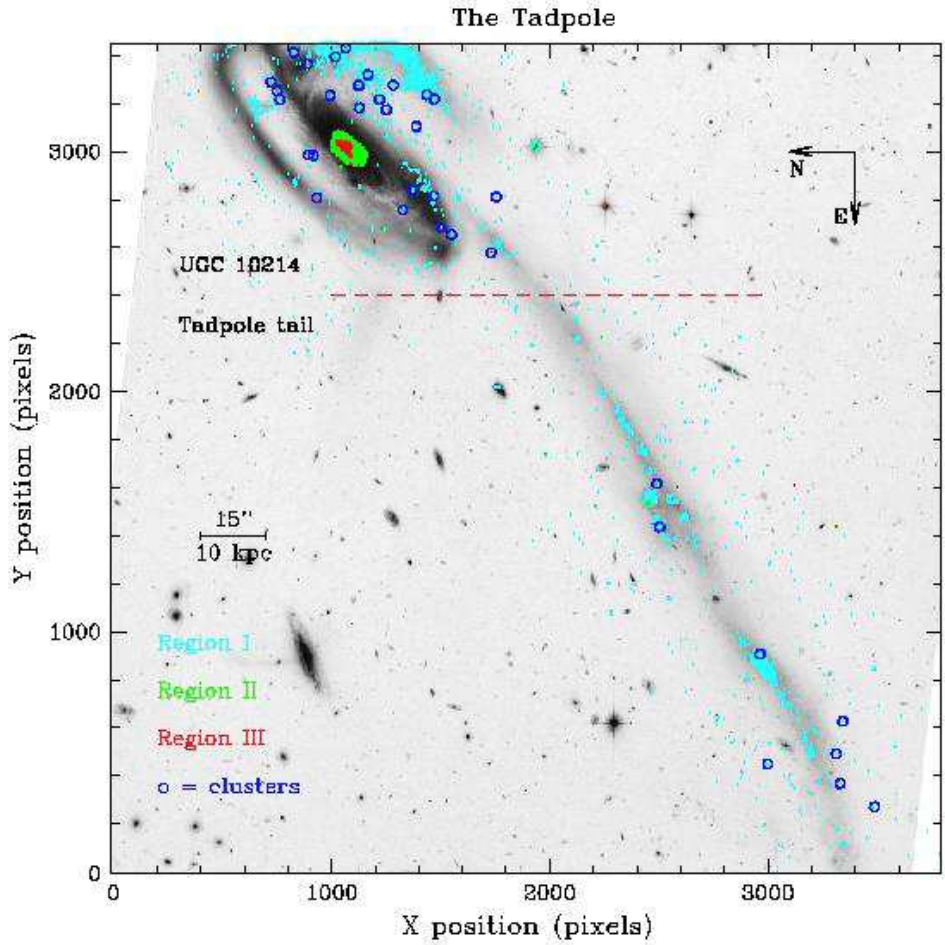,width=14cm}
\caption{\label{postadp.fig}Representation of the \astrobj{Tadpole}
galaxy on a pixel-by-pixel basis overlaid on a grey-scale rendition of
the true-colour press release image.  The specific features identified
from the CC diagram of Fig.  \ref{cctadp.fig} and the CMD of Fig. 
\ref{cmdtadp.fig} are colour coded; the areas adopted for our comparison
of the physical properties of the main galactic disk and the tidal tail
are separated by the horizontal dashed line.}
\end{figure}

In Fig.  \ref{postadp.fig} we show the physical locations in the
\astrobj{Tadpole} system corresponding to these three features.  The
blue plateau in its CC diagram, Region I, corresponds to regions in
which active star formation is currently taking place.  A large fraction
of the star clusters, indicated by circles, appears to be associated
with these areas as well. The sharply delineated bright peak in the
\astrobj{Tadpole} CMD corresponds to the core of the dominant spiral
galaxy \astrobj{UGC 10214}, a projected area occupying roughly $3.1
\times 1.2$ kpc along the galaxy's major axis, while the bright fuzzy
swarm of red pixels is located in the galactic disk underlying the sharp
bright peak in the centre region of $\sim 6.2 \times 3.1$ kpc, elongated
along the galaxy's major axis and not corrected for the galaxy's
inclination.  Although these latter pixels have redder colours than
those in the galactic centre, a close inspection of the individual
observations and of the ``true-colour'' image of \citeasnoun{F02} does
not reveal significant amounts of extinction surrounding the immediate
centre of the galaxy; these pixels are not associated with the dust lane
seen at slightly larger radii.  Thus, we conclude that the central galactic disk
stellar population is intrinsically redder than the population in its
very core.  This may indicate stronger present or more recent star
formation in the core \citeaffixed{W93}{see e.g.  \astrobj{NGC 7252};},
or perhaps low-level nuclear (Seyfert or LINER) activity, such as seen
in a number of early-type spiral galaxies with blue cores
\citeaffixed{W02}{e.g.}.  Near-infrared photometry in an $8''$ aperture
centred on the galactic centre \cite{J84} classifies the galactic centre
as relatively ``normal'', without the $(K-L)$ excess expected from a
recent burst of star formation. 

Finally, we compared the distributions and densities in colour-colour
and colour-magnitude space of the pixels originating in the tidal tail
and the galactic disk itself (see Fig.  \ref{postadp.fig}).  While the
colours of the tidal tail and of the main galactic disk occupy very
similar areas in the CC diagram, the pixels in the tidal tail are {\it
dominated} by ``Region I'' colours: $\sim 60$\% of the pixels in the
tidal tail correspond to colours in Region I.  They contribute $\sim 70,
55,$ and 40\% of the total flux in the F475W, F606W, and F814W
passbands, respectively.  (Only a small fraction of the light from the
main disk of the galaxy corresponds to ``Region I''-type colours.) The
colours of the remaining pixels coincide with the colours of the main
galactic disk. 

\subsubsection{Unraveling \astrobj{the Mice}}

\begin{figure}
\psfig{figure=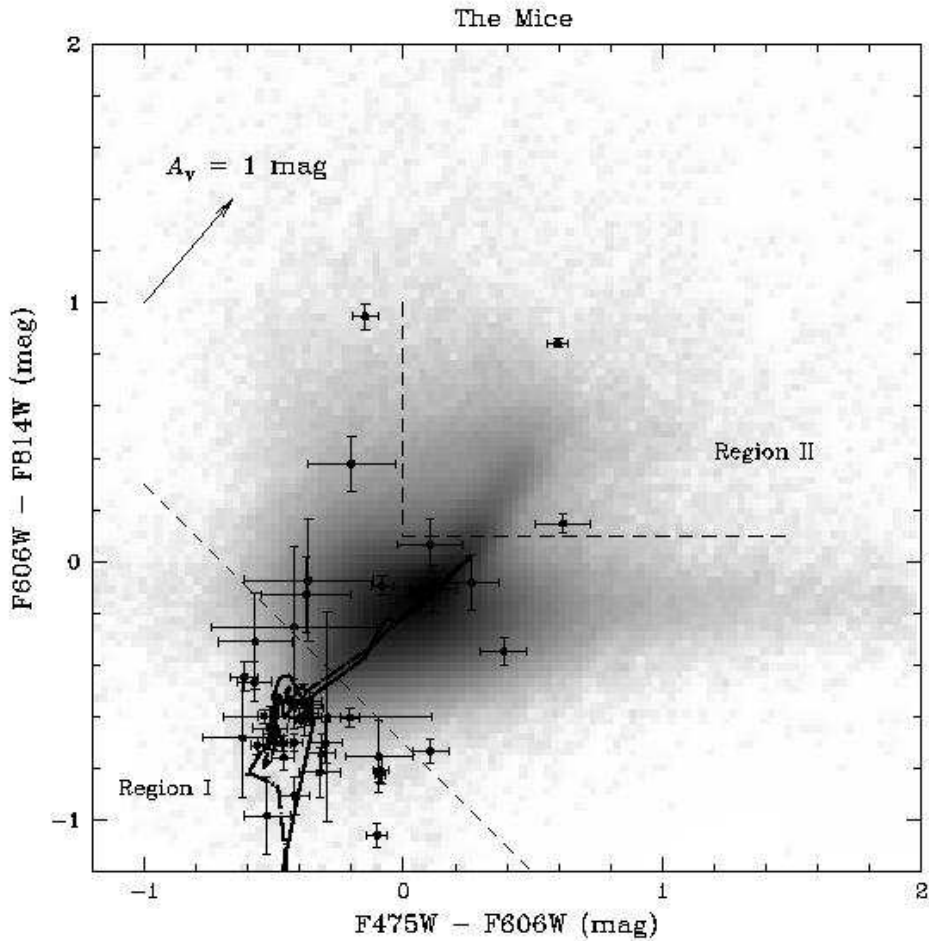,width=14cm}
\caption{\label{ccmice.fig}Pixel CC diagram for \astrobj{the Mice}. 
Figure coding is as in Fig.  \ref{cctadp.fig}.}
\end{figure}

Although the drop in density towards bluer colours in the CC diagram of
\astrobj{the Mice} (see Fig.  \ref{ccmice.fig}) is not as sharp as for
the \astrobj{Tadpole}, we see that the star clusters are predominantly
found in the same region in colour-colour space as before.  Therefore,
we adopt the same definition as for the \astrobj{Tadpole} for Region I
in \astrobj{the Mice}.  The difference between \astrobj{the Mice} and
the \astrobj{Tadpole} in the sense that we observe a more gradual
decrease in density towards bluer colours in \astrobj{the Mice} than in
the \astrobj{Tadpole} indicates most likely a real difference in the SFH
of both systems.  In the \astrobj{Tadpole}, the interaction has affected
only small regions of the system because of the relatively minor
gravitational disturbance of the lower-mass galaxy on the higher-mass
one (see Section \ref{\astrobj{Tadpole}.sect}), while the active star
formation is more widespread throughout both galaxies and the tidal
tails of the Mice. 

\begin{figure}
\psfig{figure=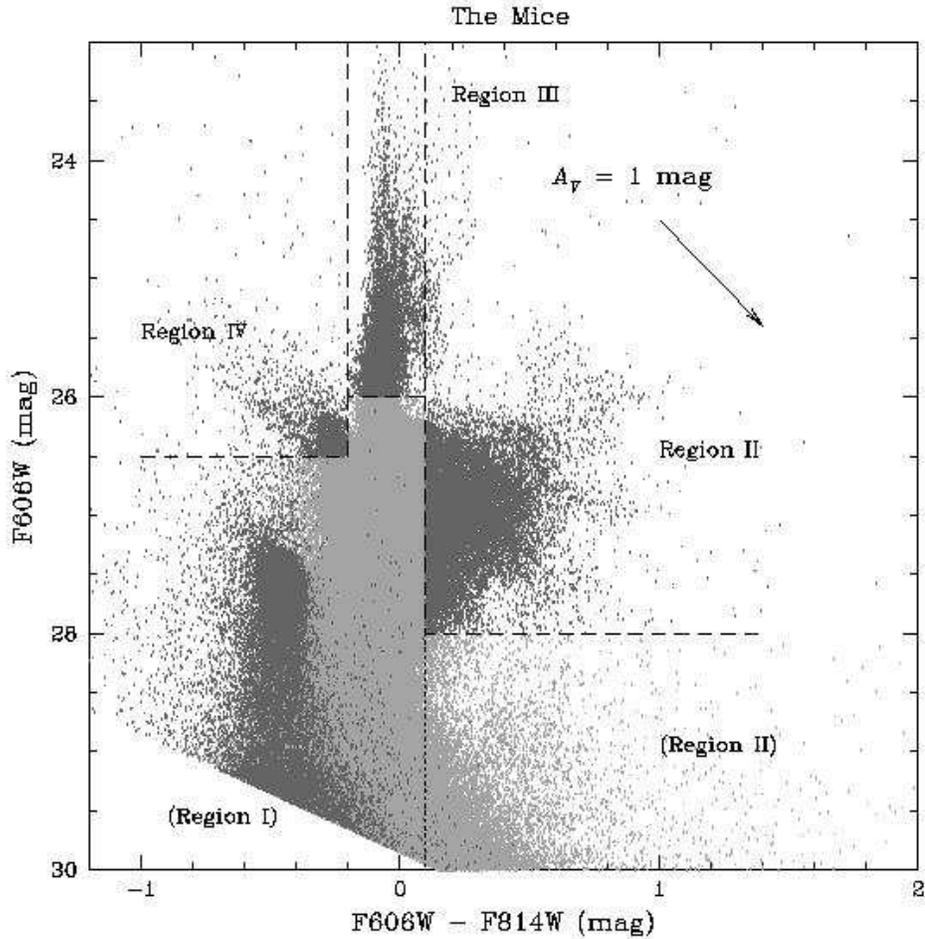,width=14cm}
\caption{\label{cmdmice.fig}Pixel CMD for \astrobj{the Mice}.  Figure
coding is as in Fig.  \ref{cmdtadp.fig}. The definition of Region II
based on the CC diagram of Fig. \ref{ccmice.fig} includes all pixels
redder than the vertical demarcation at $({\rm F606W}-{\rm F814W}) =
0.1$; based on this CMD, however, we further restricted this region to
pixels brighter than F606W = 28 mag.}
\end{figure}

The \astrobj{Mice} CC diagram shows a red extension, roughly
characterised by $({\rm F475W}-{\rm F606W}) \ge 0.0$ mag and $({\rm
F606W}-{\rm F814W}) \ge 0.2$ mag.  Further verification of this feature
in the CMD (Fig.  \ref{cmdmice.fig}) shows that we also need to restrict
the pixels in this feature, Region II, to obey F606W $\le 28.0$ mag, to
avoid contamination by low-S/N pixels and background noise. 

The ``plume'' of enhanced density in Fig.  \ref{ccmice.fig} at
$({\rm F606W}-{\rm F814W}) \sim -0.2$ mag is an artificial feature
caused by the lowest-S/N pixels in the F475W passband. 

We define two further regions of interest in \astrobj{the Mice}: Region
III consists of the bright peak with pixel values brighter than F606W =
26.0 mag, while it appears that a further plume in the CMD, starting at
the faint limit of the bright peak, but with bluer colours, might be a
physically distinct population of pixels.  We define Region IV to be
composed of pixels with F606W $\le 26.5$ mag, and $({\rm F606W}-{\rm
F814W}) \le -0.2$ mag.  Again, all of these features occupy similarly
well-defined loci in the other CMDs that can be constructed from the
available observations. 

\begin{figure}
\psfig{figure=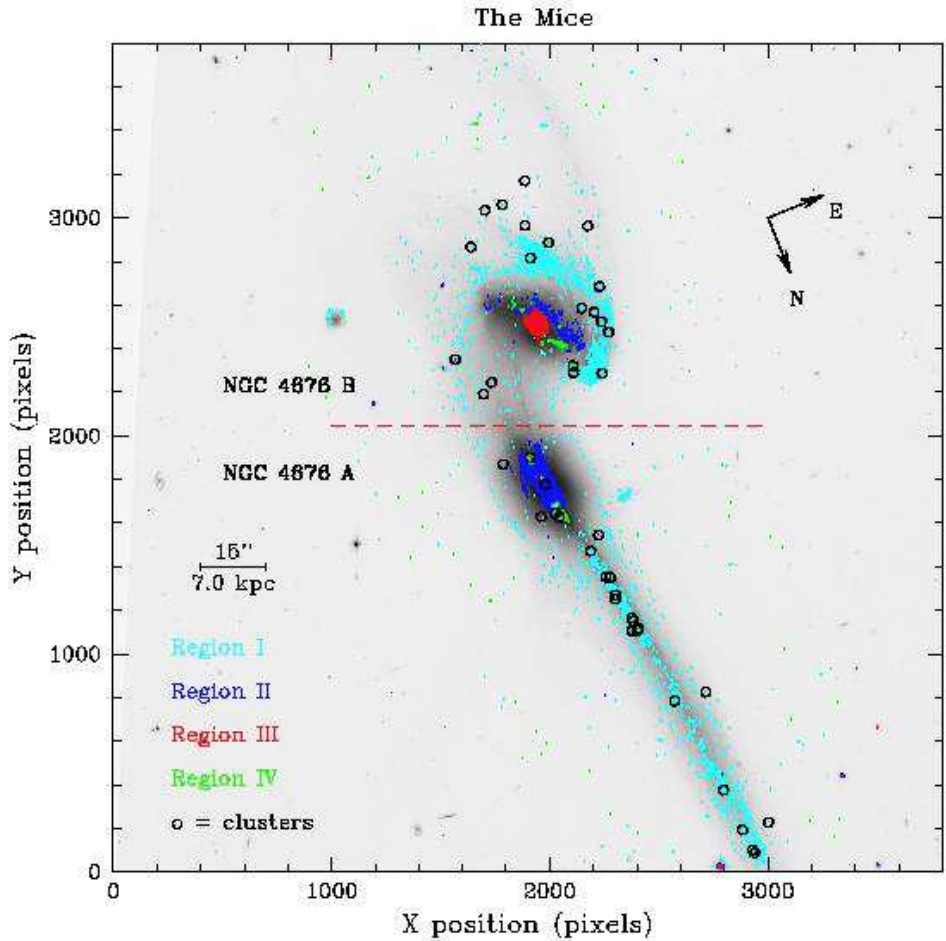,width=14cm}
\caption{\label{posmice.fig}Representation of the \astrobj{Mice} system
on a pixel-by-pixel basis.  The specific features identified from the CC
diagram of Fig.  \ref{ccmice.fig} and the CMD of Fig.  \ref{cmdmice.fig}
are colour coded, and overlaid on a grey-scale rendition of the
true-colour press release image; the areas adopted for our comparison of
the physical properties of both galaxies are separated by the horizontal
dashed line.}
\end{figure}

In Fig.  \ref{posmice.fig}, we show the correspondence between the
features in the CC diagram and the CMDs of \astrobj{the Mice} and their
physical locations in the system itself.  Again, we see that all of the
features we identified coincide with physically distinct regions in the
galaxies.  Region I again coincides with regions of active star
formation, and a fairly good match with the locations of the star
clusters is observed once more.  Region II is associated with areas of
heavy extinction. 

Although we did not define a similar region as Region II in the CC
diagram of the \astrobj{Tadpole} galaxy, this is merely due to the
sharper demarcation of Region II in \astrobj{the Mice}, and not to the
absence nor even the lower density of such red pixels in the
\astrobj{Tadpole} system.  The sharper demarcation in \astrobj{the Mice}
indicates that the dust and stellar populations are less well mixed in
this system than in the \astrobj{Tadpole} (or that dust is more
important in \astrobj{the Mice}).  We attribute this to the difference
in viewing angle between the two systems.  The dust lanes are most
clearly seen and most sharply defined in the edge-on view of
\astrobj{NGC 4676A}, while the dust is seen through an upper layer of
stars in the \astrobj{Tadpole} galaxy, which hence appears to be better
mixed in projection. 

The bright peak of Region III is clearly identified as the population of
pixels originating from the centre of the southern galaxy, \astrobj{NGC
4676B}.  The width of the bright peak in the CMD of \astrobj{the Mice}
is smaller than that in the \astrobj{Tadpole} CMD.  It therefore appears
that the contrast between the blue core and the redder central region in
the \astrobj{Tadpole} is of real significance.

The slightly fainter but bluer feature of Region IV is associated with
central star forming regions in both galaxies in the system.  Both
\astrobj{NGC 4676A} and \astrobj{NGC 4676B} are located in the region
occupied by the ``normal'' galaxies in the near-infrared $(K-L)$ vs. 
$(H-K)$ diagram of \citeasnoun{J84}, without the $(K-L)$ excess
indicative of recent nuclear star formation.  This is confirmed by the
relatively inconspicuous and red CMD features associated with the
galactic centres, and the low far-infrared luminosity of the system
\cite{H95,HvG96}. 

A comparison between the CMDs and CC diagrams of the individual galaxies
of \astrobj{the Mice} does not reveal significant differences that
cannot be explained by the different viewing angles.  For instance,
Region II is not as densely populated in the CC diagram of \astrobj{NGC
4676B} as in that of \astrobj{NGC 4676A}, but this is likely due to the
more face-on orientation of the former galaxy.  Region I is
approximately similarly well populated in both CC diagrams; $\sim (1.5 -
2.5)$\% of the total number of pixels with F475W fluxes in excess of the
$4 \sigma_{\rm sky}$ level are found in Region I.  This indicates that
the burst of star formation triggered by the encounter is similar in
both galaxies and has affected only relatively small areas in either
galaxy, in particular in their tidal tails and outer spiral arms. 

\section{Implications for star and cluster formation scenarios}
\label{implications.sect}

\subsection{Dynamical time-scales and cluster formation in \astrobj{the
Mice}} 
\label{mice.sect}

Starting with \citeasnoun{TT72}'s seminal paper based on a small number
of test particles, sophisticated {\it N}-body simulations have been
employed to generate an interaction morphology resembling that observed
for \astrobj{the Mice} \citeaffixed{M93,GS93,B96,SR98}{e.g.}.  The
current state of the art simulations seem to converge to agreement on
the interaction geometry, inclination angles of the individual galaxies,
parabolic orbital parameters, and a mass ratio of the two galaxies of
roughly unity \citeaffixed{B96,SR98}{e.g.} to 1:2 for \astrobj{NGC
4676A} : \astrobj{NGC 4676B} \cite{H95,HvG96} [from H{\sc i} rotation
curves].  While the morphological features are fairly well reproduced in
most respects, matching the kinematic properties of the entire system
including the tidal tails requires the presence of extended massive dark
matter haloes surrounding both galaxies \cite{SR98}, but see
\citeasnoun{GS93} for an opposing view. 

The system's morphology suggests that \astrobj{the Mice} represents a
galaxy merger in its very early stages \citeaffixed{T77}{e.g.}; current
best estimates for the time-scale since pericentre are $\sim (160-180)
\times 10^6$ yr \cite{M93,B96}.  In order to facilitate a comparison of
the relevant time-scales, in Figs.  \ref{cctadp.fig} and
\ref{ccmice.fig} we have overplotted the model predictions for the
evolution of SSPs, based on the SSP models of \citeasnoun{S02}, properly
folded through the {\sl HST}/ACS filter response curves and taken
account of the better-than-expected in-flight performance (Gilliland,
priv.  comm.).  We added the age-dependent contributions of an
exhaustive set of gaseous emission lines and of gas continuum emission
to the \citeasnoun{S02} models.  Emission-line contributions to optical
broad-band fluxes are very important during the first $\simeq 3 \times
10^7$ yr of evolution for subsolar metallicities ($Z = 0.004 = 0.2
Z_\odot$).  For solar metallicity SSPs, their contribution is slightly
lower but still important for $t \lesssim 1.2 \times 10^7$ yr
\cite{A02}.  We assume roughly solar metallicity for the star clusters:
for star formation associated with any reasonable interaction scenario
in the present-day Universe one can expect metallicities in the range
$(0.5 - 1.0) Z_\odot$ \cite{FG94}. 

Due to the almost parallel age and extinction vectors, we cannot
determine the ages of the star clusters unambiguously, nor correct their
magnitude measurements for the effects of extinction.  However, Fig. 
\ref{ccmice.fig} shows that the majority of the bright clusters in
\astrobj{the Mice} are located in Region I, with only a small spread
along the extinction and age vectors.  The other, redder clusters could
be either part of an older cluster population, or more heavily affected
by extinction.  If we assume that for the blue clusters in Region I the
combination of foreground and internal extinction is small or
negligible, this allows us to place a lower limit on their age. 
However, such a lower limit is highly uncertain due to two further
complications: {\it (i)} the majority of the bluest star clusters are
located in the region of the CC diagram where the models are
multi-valued; and {\it (ii)} the resulting age estimates are a strong
function of metallicity for such young ages, even for a close-to-solar
metallicity range.  Nevertheless, the available data for the Mice imply
a minimum age for the majority of the blue clusters of $(1.0 \pm 0.2)
\times 10^8$ yr for solar metallicity, or $t_{\rm min} \sim 2 \times
10^8$ yr for $Z = 0.4 \, Z_\odot$.  (Note that because of the good match
of our models to the observed colours we can recover all of parameter
space by mixing various amounts of young and old[er] stellar populations
with extinction.  This implies that our basic assumption that we can use
pixel CMDs and CC diagrams as composites of SSPs is a reasonable
approximation to reality.)

We are now in a good position to comment on the importance of star
cluster formation in the context of the tidally-induced violent star
formation in the general field population.  A large fraction of the
bluest clusters coincide spatially with the blue field population in the
spine of the \astrobj{NGC 4676A} tidal tail and the outer spiral arm of
\astrobj{NGC 4676B} (see Fig.  \ref{posmice.fig}).  H$\alpha$
observations of \astrobj{the Mice} \cite{S74,M93,H95,HvG96,SR98},
indicative of recent star formation, are consistent with a tidal origin. 
A detailed comparison of Fig.  \ref{posmice.fig} with the H$\alpha$ map
in \citeasnoun{H95} shows remarkably good agreement between the
H$\alpha$-bright condensations and features within the tidal tails and
the outer spiral arms of \astrobj{NGC 4676B}, and the spatial
distribution of the bluest pixels (Regions I and IV) in \astrobj{the
Mice} (Figs.  \ref{ccmice.fig} and \ref{cmdmice.fig}). 

We can, in fact, place a strong lower limit on the age of the
H$\alpha$-bright regions in \astrobj{the Mice}.  \citeasnoun{S74}
concluded, from his H$\alpha$+[N{\sc ii}] spectroscopy, that all of the
emission from \astrobj{NGC 4676A}, including the entire length of the
tidal tail, and from at least part of \astrobj{NGC 4676B} is A-star
dominated.  The most likely origin for such an A-type spectrum is rapid,
widespread star formation that effectively ceased $\gtrsim 5 \times
10^7$ yr ago, so that earlier-type O and B stars would have had
sufficient time to evolve off the main sequence.  This is consistent
with the {\it current} star formation rates derived by \citeasnoun{SR98}
in both \astrobj{NGC 4676A} and its tidal tail, which are not unlike the
rates found in normal spiral galaxies.  This implies, therefore, that
these parts of the interacting system are currently not undergoing
enhanced star formation due to the encounter, but did so $> 5 \times
10^7$ yr ago. 

Secondly, \citeasnoun{E93} and \citeasnoun{SR98} suggest that the actual
mechanism responsible for the star formation in the H$\alpha$
condensations is driven by gravitational instabilities in the gas caused
by the encounter.  They calculate that the time-scale for developing
such instabilities is on the order of a few times $10^7$ yr, which
places a similar lower limit on the age of these compact objects. 
Therefore, the main event triggering the cluster formation must have
occurred at least $\sim (1.5-2.0) \times 10^8$ yr ago, depending on the
precise cluster metallicities.

We conclude, therefore, that Regions I and IV in the CC diagram of
\astrobj{the Mice} correspond to active and recent star formation
induced by the tidal interaction, while the star clusters found
scattered throughout these star-forming regions are {\it also}
consistent with having been formed on similar time-scales. 

The fractional contribution of the star clusters to the total flux in
Region I in \astrobj{the Mice} is $\sim 40$\%, independent of
wavelength.  Because of our selection bias towards brighter star
clusters, this is in fact a lower limit to the total cluster
contribution.  Despite the fact that this region is not as sharply
defined in \astrobj{the Mice} as in the \astrobj{Tadpole} system, this
clearly shows that star cluster formation is a major mode of star
formation in galactic encounters. 

Some 40\% of the star cluster candidates identified in Section
\ref{clusters.sect} coincide spatially (in projection) with the
H$\alpha$-bright condensations, which therefore might be either young
star clusters, knots of active star formation, giant H{\sc ii} complexes
\cite{H95,HvG96,SR98}, or due to projection effects.  Most of the
objects are unresolved by the ACS PSF, and therefore have diameters of
$\lesssim 35$ pc. 

If we assume that the ongoing tidal interactions between \astrobj{NGC
4676A} and \astrobj{NGC 4676B} have indeed induced the formation of the
observed bright blue star clusters in the past $\sim 2 \times 10^8$ yr,
we can estimate the cluster masses and compare those to the masses of
star clusters in other young cluster systems.  The bright blue star
clusters in \astrobj{the Mice} exhibit a peak in their brightness
distribution at F606W $\sim 22$ mag; combined with the mass-to-light
($M/L$) ratio in the F606W passband at an age of $\lesssim 2 \times
10^8$ yr, $M/L_{606} \lesssim 0.2$, we estimate a characteristic
photometric mass for these clusters of $\lesssim 3 \times 10^6 M_\odot$. 

We emphasize that we are only sampling the bright wing of the CLF, and
therefore the high-mass wing of the cluster mass function.  Independent
dynamical mass estimates are available only for a few of the most
luminous SSCs, and are approximately $10^6 M_\odot$
\cite{HF96a,HF96b,SG01}.  The masses of Galactic globular clusters are
typically in the range $10^4 - 3 \times 10^6 M_\odot$
\citeaffixed{M91,PM93}{e.g.}.  Thus, the approximate photometric mass
estimate obtained for the median of the bright blue clusters in
\astrobj{the Mice} is consistent with this being the bright wing of a
``normal'' SSC or globular cluster progenitor population. 

\subsection{Star and cluster formation in the \astrobj{Tadpole} galaxy}
\label{\astrobj{Tadpole}.sect}

The \astrobj{Tadpole} system is an example of a galaxy encounter between
two unequal-mass galaxies at a later stage.  The system is composed of a
mildly disturbed SBc galaxy and a long, low-S/N tidal tail.  The latter
shows a number of bright, blue condensations, while active star
formation is also apparent on the inside of the western spiral arm (Fig. 
\ref{postadp.fig}).  \citeasnoun{F02} attribute this region of enhanced
star formation to the remnant of a low-mass galaxy that caused the
disturbed appearance of the system. 

As for \astrobj{the Mice}, in Fig.  \ref{cctadp.fig} we have overplotted
the model predictions for the evolution of SSPs.  We see that the
majority of the star clusters are located in a very similar region as
the clusters in \astrobj{the Mice}, again with a relatively small spread
along the age/extinction vector.  In view of the formation constraints
imposed by the time-scale to develop gravitational instabilities, we
conclude that the clusters we identified in the \astrobj{Tadpole} (Table
\ref{tadpclus.tab}) must also have been formed very recently, $\sim (1.5
- 2.0) \times 10^8$ yr ago, and represent the high-mass wing of a
similar mass distribution as in \astrobj{the Mice}.  This minimum age
estimate is consistent with their spatial distribution throughout the
galactic disk and tidal tail, which matches the actively star forming
regions defined by the pixels in Region I reasonably well.  The clusters
located in Region I of the \astrobj{Tadpole} CMD contribute $\gtrsim
35$\% to the total flux of all pixels in this region, again independent
of passband for the filter combination used here. 

It is rather remarkable that the interaction of a massive SBc galaxy and
a low-mass companion caused such a strong deformation of the massive
galaxy and also very pronounced star {\it cluster} formation.  We can
obtain a rough, independent estimate of the dynamical age of the
interaction in the \astrobj{Tadpole} interacting system by comparing the
length of its tail ($\sim 110$ kpc) to the stellar velocity dispersion
expected in the undisturbed progenitor galaxy.  In a typical large
spiral galaxy, velocity dispersions of $\sim (150-300)$ km s$^{-1}$ are
expected \cite{Tr02}, so that the dynamical age since the beginning of
the interaction resulting from this impulse approximation is $t_{\rm
dyn} \sim (4-8) \times 10^8$ yr.  This is remarkably close to the ages
of the star clusters estimated from their loci in the CC diagram. 

\section{Summary and Conclusions}
\label{summary.sect}

We have utilised optical pixel-by-pixel CMDs and CC diagrams to explore
and deduce the star and star cluster formation histories of \astrobj{the
Mice} and the \astrobj{Tadpole} interacting galaxies.  Despite their
considerable distance of $\gtrsim 100$ Mpc, this technique now offers
promising prospects to study the local, dominant stellar populations out
to greater distances than ever before thanks to the capabilities of the
Advanced Camera for Surveys onboard {\sl HST}.  The analysis performed
in this paper is based on a subset of the ACS Early Release
Observations, obtained from the {\sl HST} data archive.  One of the main
advantages of ACS/WFC observations is that the PSF is (marginally)
undersampled, so that each pixel corresponds to a spatially independent
measurement of the underlying composite stellar population. 

In both interacting systems we find some 40 bright young star clusters
($20 \lesssim {\rm F606W (mag)} \lesssim 25$, with diameters $\lesssim
35$ pc and a characteristic mass of $\sim 3 \times 10^6 M_\odot$), which
are spatially coincident with blue regions of active star formation in
the tidal tails and spiral arms of our target galaxies.  Despite the
complications caused by the age--extinction degeneracy in the
combination of the F475W, F606W and F814W optical passbands, we estimate
that the main event triggering the formation of these clusters occurred
$\sim (1.5 - 2.0) \times 10^8$ yr ago.  It is therefore likely that the
gravitational interactions responsible for the disturbed appearances of
our target galaxies have induced not only enhanced star formation, but
also star cluster formation.  We argue that star cluster formation is a
major mode of star formation in galactic interactions, with $\gtrsim
35$\% of the active star formation in encounters occurring in star
clusters.  This is the first time that young star clusters have been
detected along the tidal tails in interacting galaxies. 

We find that specific features and plumes in the CMDs and CC diagrams of
both interacting systems correspond to sharply demarcated regions in the
systems themselves.  In the \astrobj{Tadpole} system, we detect a
hitherto unknown very bright blue core superimposed on a redder central
area, which we speculate may be related to some form of nuclear activity
in the main disk galaxy of this system, \astrobj{UGC 10214}.  The tidal
tail of the system is dominated by blue star forming regions, which
occupy some 60\% of the total area covered by the tail; they contribute
$\sim 70$\% of the total flux in the F475W passband (decreasing to $\sim
40$\% in F814W).  The remaining pixels in the tail have colours
consistent with those of the main disk. 

The \astrobj{Mice} interacting galaxies are an example of an
approximately equal-mass encounter.  We identify regions of pronounced
star formation in the tidal tails, outer spiral arms, and near the
galactic centres.  We conclude that the tidally triggered burst of star
formation is of similar strength in both galaxies, but it has affected
only relatively small spatially coherent areas. 

Finally, the analysis in this paper has shown that the technique of
using both CMDs and CC diagrams on a pixel-by-pixel basis, combined with
measurements of individual, distinct compact objects, is a very powerful
tool to unravel the complex star formation histories and their
associated star cluster formation in interacting galaxies. 

\section*{Acknowledgements} JTL and MCMH acknowledge PPARC funding to
attend the 2002 PPARC/ Cambridge International Undergraduate Summer
School.  We thank Paul Eskridge, Max Mutchler, Ronald Gilliland, Rolf
Jansen, and Richard Sword for scientific or graphics suggestions and
support.  This research has made use of NASA's Astrophysics Data System
Abstract Service.  The Interactive Data Language ({\sc idl}) is licensed
by Research Systems Inc., of Boulder, CO, USA.

\end{document}